\documentclass[%
 pra,
 amsmath,amssymb,
reprint,%
superscriptaddress,
floatfix,
]{revtex4-2}

\usepackage{graphicx,float}% Include figure files
\usepackage{dcolumn}% Align table columns on decimal point
\usepackage{bm,braket}% bold math
\usepackage{amsfonts}
\usepackage{dsfont}
\usepackage[utf8]{inputenc}
\usepackage[T1]{fontenc}
\usepackage{dsfont,siunitx,booktabs}
\usepackage{hyperref}
\usepackage{soul}
\usepackage{amsmath}
\begin{document}

%\preprint{AIP/123-QED}

\title{Tuning the entanglement growth in matrix-product-state evolution of quantum systems by nonunitary similarity transformations}
% Force line breaks with \\

\author{Hanggai Nuomin}%

\affiliation{ 
Department of Chemistry, Duke University, Durham, North Carolina 27708, United States
%\\This line break forced with \textbackslash\textbackslash
}
\author{Feng-feng Song}%
%\email{sff18@mails.tsinghua.edu.cn}
\affiliation{ 
Department of Physics, Tsinghua University, Beijing 100084, China
}
\author{Peng Zhang}%
\email{peng.zhang@duke.edu.}
\affiliation{ 
	Department of Chemistry, Duke University, Durham, North Carolina 27708, United States
	%\\This line break forced with \textbackslash\textbackslash
}%
\author{David N. Beratan}%
\email{david.beratan@duke.edu.}
\affiliation{ 
	Department of Chemistry, Duke University,
	Durham, North Carolina 27708, United States
}
\affiliation{ 
	Department of Physics, Duke University,
	Durham, North Carolina 27708, United States
}
\affiliation{ 
	Department of
	Biochemistry, Duke University,
	Durham, North Carolina 27710, United States
}

\date{\today}% It is always \today, today,

\begin{abstract}
The possibility of using similarity transformations to alter dynamical entanglement growth in matrix-product-state simulations of quantum systems is explored. By appropriately choosing the similarity transformation, the entanglement growth rate is suppressed, improving the efficiency of numerical simulations of quantum systems. The transformation can be applied to general quantum-many-body systems.
\end{abstract}
\maketitle

\section{Introduction}
Open-quantum-system dynamics can be efficiently simulated using tensor-network techniques \cite{baxter1968dimers,affleck1987rigorous,white1992density,white1993density,white2004real,daley2004time,schollwock2011density,schroder2016simulating} (e.g. matrix product states) in low-entanglement regimes. These regimes often have weak system-bath couplings and low temperatures (if no phase transition happens). 
However, in the high-entanglement regimes, MPS simulations of quantum many-body systems remain challenging and are being actively studied \cite{chin2010exact,prior2010efficient,karrasch2012finite,rams2020breaking,shi2018ultrastrong,schroder2016simulating,liu2021improved,kohler2021efficient}. Most of the proposed methods used global basis transformations to reduce entanglement in the MPS of quantum systems, since local on-site unitary transformations can not change the entanglement of a quantum state.

In this letter, we introduce a local non-unitary similarity transformation for open-system Hamiltonians and examine its impact on the entanglement growth rate during the real-time simulations of open quantum systems.
The transformation produces a family of Hamiltonians characterized by a continuous parameter that controls the entanglement growth rate. By tuning the parameter, one can suppress the entanglement growth in matrix product states (MPS) during real-time dynamical simulations of highly entangled quantum systems. This method is general and can be applied to almost any Hamiltonians, including open-quantum-system Hamiltonians with non-harmonic baths or non-linear system-bath couplings. It sheds light on reducing entanglement in the MPS simulations of quantum many-body system. The effect of non-unitary similarity transformations on the simulation efficiency has been explored extensively in the electronic-structure literature (see, for example, Refs \cite{wahlen2015lie,baiardi2020transcorrelated,dobrautz2019compact,chan2005density}), but is much less explored for quantum dynamics. This letter investigates the use of similarity transformations to tune the entanglement growth in \emph{time-dependent} simulations of open quantum system dynamics. 

\section{THEORY AND METHODOLOGY}
\label{sec:theory}
Given a general  Hamiltonian of an open quantum system coupled to a bath
\begin{align}
{H}={H}_{s} + {H}_{sb} + {H}_{b}
\label{eq:open-Hamiltonian}
\end{align}
where $H_{b}$ is a collection of bosons (interacting or non-interacting) and ${H}_{sb}$ is bi-linear couplings between the system and the bosons , it is possible to obtain a new Hamiltonian $\mathcal{H}$ with a similarity transformation:  $\mathcal{H}=e^{\hat{S}}{H}e^{-\hat{S}}$. Here, we specifically choose $\hat{S}$ to be Hermitian, in contrast to the usual choice that $\hat{S}$ is anti-Hermitian. Thus, $e^{\hat{S}}$ is non-unitary: $(e^{\hat{S}})^\dagger = e^{\hat{S}}$. In the similarity-transformed frame, with the help of an evolution operator $\mathcal{U}_t = e^{-i\mathcal{H}t}$, the time evolution of the expectation value (which shall be transformation-invariant) for an observable $\hat{A}$ is given by
\begin{align}
    \braket{\hat{A}}(t) = \braket{\psi_0|e^{\hat{S}}\mathcal{U}^\dagger_t e^{-\hat{S}} A e^{-\hat{S}} \mathcal{U}_t e^{\hat{S}}|\psi_0}.
\end{align}
We can recognize $e^{-S} \mathcal{U}_t e^{S}\ket{\psi_0}$ is the state vector at time $t$ in the original frame (i.e., without the similarity transformation), and $\mathcal{U}_t e^{S}\ket{\psi_0}$ is the state vector at time $t$ in the transformed frame.

A judiciously chosen $\hat{S}$ can alter the nature of the Hamiltonian and favors the numerical simulations of the Hamiltonian in Eq.~\eqref{eq:open-Hamiltonian}. We will use the zero-bias spin-boson model as an example:
\begin{align}
	\hat{H} = \Delta\hat{\sigma}_x + \sum_{n}c_n \hat{A} \otimes (\hat{a}^\dagger_n + \hat{a}_n) + \sum_{n}\omega_c \hat{a}^\dagger_n\hat{a}_n.
	\label{eq:actual-Hamiltonian}
\end{align}
where $\Delta$ is the coupling between two spin states $\ket{\uparrow}$ and $\ket{\downarrow}$, and $c_n$ and $\omega_n$ are the coupling strength and the vibration frequency associated to the $n$-th boson. In the following, we use $\hat{A}=\hat{\sigma}_z$.
The simplest possible choice of $\hat{S}$ is $\hat{S}=\beta\hat{\sigma}_z$ with $\beta$ being a constant. Other forms of $\hat{S}$ are possible (e.g. $\beta\protect\sum_{n}\hat{a}^\dagger_n \hat{a}_n$). We choose $\beta\hat{\sigma}_z$ as a proof of principle. The transformed Hamiltonian $\mathcal{H}$ is then:
\begin{align}
	\mathcal{H}(\beta) = e^{\beta\hat{\sigma}_z}\hat{\sigma}_x e^{-\beta\hat{\sigma}_z} + \sum_{n}c_n \hat{\sigma}_z \otimes (\hat{a}^\dagger_n + \hat{a}_n) + \sum_{n}\omega_n \hat{a}^\dagger_n\hat{a}_n.
	\label{eq:transformed-Hamiltonian}
\end{align}
To better understand the effect of the similarity transformation, we give explicitly the matrix elements of $e^{\beta\hat{\sigma}_z}\sigma_xe^{-\beta\hat{\sigma}_z}$:
\begin{align}
	e^{\beta\hat{\sigma}_z}\sigma_xe^{-\beta\hat{\sigma}_z} = 
	\begin{pmatrix}
		0 & e^{2\beta}\\
		e^{-2\beta} & 0
	\end{pmatrix}.
	\label{eq:transformed-spin}
\end{align}
This matrix means that after the transformation, the transition from $\ket{\downarrow}$ to $\ket{\uparrow}$ is enhanced, while the reverse transition is diminished. This scenario of incommensurate transition strengths is reminiscent of the  Aubry-Andr\'e-Harper model \cite{harper1955single,aubry1980analyticity} which can be realized in optical-lattice experiments.

If the initial state of the spin is an eigenstate of $\hat{\sigma}_z$ (e.g., $\ket{\uparrow}$), which is the case for most of the open quantum system simulations, the effect of the non-unitary transformation $e^{\beta\hat{\sigma}_z}(\cdot)e^{-\beta\hat{\sigma}_z}$ is, to some extent, to freeze the spin in its initial eigenstate. The freezing effect is favorable to the matrix-product-state simulation of open quantum systems since the initial state often has the lowest entanglement. The parameter $\beta$ characterizes a family of Hamiltonians $\mathcal{H}(\beta)$ which are related to each other by the non-unitary transformation. It is expected that some of the Hamiltonians (i.e.,  some special values of $\beta$) in this family can show slower growth of entanglement during the evolution than the original Hamiltonian $\mathcal{H}(0)$.

The transformed Hamiltonian (Eq.~\eqref{eq:transformed-Hamiltonian}) describes a fictitious system which is connected to the actual system (Eq.~\eqref{eq:actual-Hamiltonian}) by the non-unitary transformation $e^{\beta\hat{\sigma}_z}(\cdot)e^{-\beta\hat{\sigma}_z}$. The dynamics of the fictitious system can be back-transformed to the dynamics of the actual system by a recovering transformation $e^{-\beta\hat{\sigma}_z}(\cdot)e^{-\beta\hat{\sigma}_z}$. The density matrix of the fictitious system $\rho_f(t)=\mathcal{U}_t e^{S}\ket{\psi_0}\bra{\psi_0}e^{S}\mathcal{U}_t^\dagger $ is related to the density matrix of the actual system $\rho(t)$ by:
\begin{align}
	\rho(t) = \frac{ e^{-\beta\hat{\sigma}_z}\rho_f(t)e^{-\beta\hat{\sigma}_z}}{\mathrm{tr}[e^{-\beta\hat{\sigma}_z}\rho_f(t)e^{-\beta\hat{\sigma}_z}]}.
	\label{eq:density-matrix-relation}
\end{align}
From this relation, it is clear that the fictitious reduced density matrix of the spin $\mathrm{tr}_B\rho_f(t)$ is still a legitimate density matrix---after a proper normalization: $\rho_f(t)\to \rho_f(t)/\mathrm{tr}(\rho_f(t))$. The entanglement of the density matrix of the fictitious system can still be defined by the singular vale decomposition of the coefficient tensor of the wave-function expansion in terms of an orthonormal basis. The entanglement of the fictitious-system wave functions has no simple relation to the entanglement of the actual wave function, in contrast to the simple equation \eqref{eq:density-matrix-relation} describing the relation for the wave functions themselves. This necessitates numerical simulations to explore the properties of entanglement growth for the Hamiltonian family $\{\mathcal{H}(\beta)\}$.  We will use the zero-bias spin-boson model to show the entanglement growth for different $H(\beta)$'s in Sec.\ref{sec:results}.

% However, the similar transformation cannot be applied to $\sigma_z$ as
% \begin{align}
% 	e^{\beta\hat{\sigma}_x}\sigma_z e^{-\beta\hat{\sigma}_x} = 
% 	U^\dagger
% 	\begin{pmatrix}
% 		0 & e^{2\beta}\\
% 		e^{-2\beta} & 0
% 	\end{pmatrix}U=
% 	\begin{pmatrix}
% 		\cosh{2\beta} & -\sinh{2\beta}\\
% 		\sinh{2\beta} & -\cosh{2\beta}
% 	\end{pmatrix},
% \end{align}
% where $U=\begin{pmatrix}
% 		1/\sqrt{2} & 1/\sqrt{2}\\
% 		1/\sqrt{2} & -1/\sqrt{2}
% 	\end{pmatrix}$.

\section{Numerical Results}
\label{sec:results}
We use the Drude spectral density $J(\omega)=\frac{\eta \omega_{c} \omega}{\omega_{c}^{2}+\omega^{2}}$ for the spin-boson model where $\eta$ and $\omega_c$ are the coupling strength and the characteristic frequency of the bath modes, respectively. To include the finite-temperature effect, the spectral density is  thermalized by multiplying a temperature-dependent factor and extending the frequency axis to negative infinity according to Refs \cite{tamascelli2019efficient,dunnett2020simulating}.  The thermalized spectral density is discretized (with maximal and minimal frequencies $\omega_\mathrm{max}$ and $-\omega_\mathrm{max}$) to generate the Hamiltonian in Eq.~\eqref{eq:actual-Hamiltonian} which is then transformed to the $\beta$-dependent Hamiltonian in Eq.~\eqref{eq:transformed-Hamiltonian}. No further basis transformation (e.g., the chain transformation \cite{prior2010efficient,chin2010exact}) is used. Thus, the topology of the Hamiltonian is a ``star'' \cite{liu2021improved} which is known to have rapid growth of entanglement, if the similarity transformation is not applied. The parameters used for the spin-boson model are present in Tab.~\ref{tab:parameters}. The parameters chosen have typical values used in the simulations of open quantum systems.
\begin{table}
	\begin{tabular}{ll}\toprule
			Parameter   & Physical Quantity                                    \\\midrule
			$\Delta$      & spin coupling \\
			$\eta = 4\Delta$\      & system-bath coupling\\
			$\omega_c = 1\Delta$ & characteristic bath frequency\\
			$k_BT = 2\Delta$ & bath temperature \\
			$\pm\omega_\mathrm{max}=\pm12.7324\Delta$ & frequency cutoffs in discretization\\
			\bottomrule
	\end{tabular}
	\caption{The reduced Planck constant $\hbar$ is set to be $1$. $\eta$, $\omega$, $T$ and $\omega_\mathrm{max}$ are in the unit of $\Delta$ (spin coupling). $\pm\omega_\mathrm{max}$ are the upper and lower limit used in the discretization of the Drude spectral density.}
\label{tab:parameters}
\end{table}
The initial state of the system and the bath is set to be $\ket{\uparrow}\otimes\ket{0,0,\ldots}$. The second-order time-evolving-block-decimation (TEBD2) method \cite{vidal2003efficient} with swap gates \cite{stoudenmire2010minimally} is used to propagate the wave function. It is worth noting that the evolution operators used in TEBD2 can be non-unitary since the Hamiltonian is non-unitary itself. The non-unitary evolution operators destroy the canonical form of a MPS. To restore the canonical form and obtain the correct singular values, we perform an canonicalization procedure after each evolution step \cite{vidal2003efficient}.

\begin{figure}
	\centering
	\includegraphics{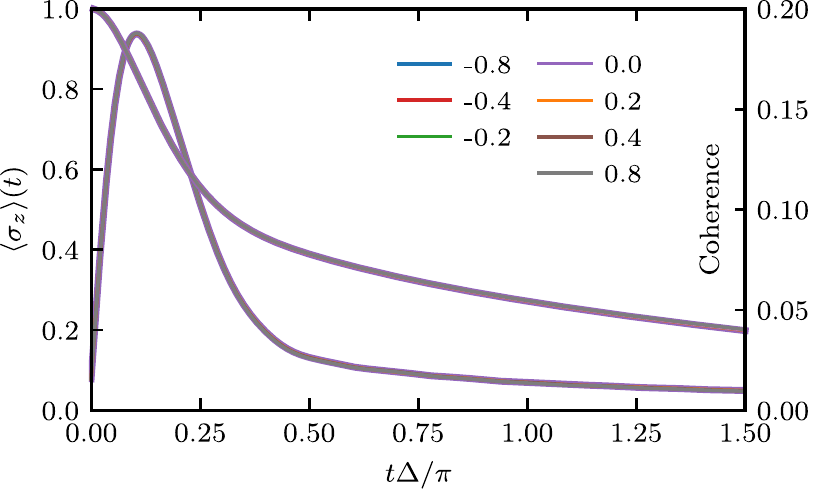}
	\caption{The dynamics of the polarization $\braket{\sigma_z}$ and the real part of non-diagonl elements of $\rho(t)$ with different $\beta$'s. The threshold for singular values is $10^{-5}$. The time step is $0.005$. If $\beta$'s larger than $0.8$ (or smaller than $-0.8$) are used, the dynamics start to show deviations from the exact line ($\beta=0$).} 
	\label{fig:pop}
\end{figure}
Fig.~\ref{fig:pop} shows the time-dependent polarization $\braket{\hat{\sigma}_z}$ obtained by the recovering equation Eq.~\eqref{eq:density-matrix-relation}. The results generated from the fictitious systems using the recovering equation agree well with the actual dynamics ($\beta=0$, the thick line), showing the correctness of the similarity-transformation formalism. The lines with $\beta>0.8$ and $\beta<-0.8$ are not showing in Fig.~\ref{fig:pop} since they start to deviate from the correct dynamics. This deviation suggests that $\beta$ could not be too large in magnitude, otherwise the accuracy of results is deteriorated.

\begin{figure}
	\includegraphics{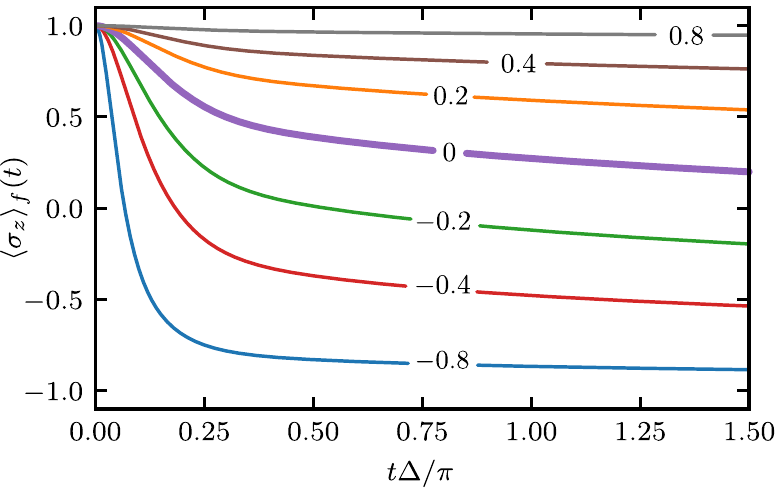}
	\caption{The dynamics of $\mathrm{tr}[\sigma_z\rho_f(t)]$ without the recovery (Eq.~\eqref{eq:density-matrix-relation}). The numbers in the middles of lines are different $\beta$'s. The population transfer from $\ket{\uparrow}$ to $\ket{\downarrow}$ is slowed down with $\beta>0$, while negative $\beta$'s accelerate the transfer. The thick line is the population dynamics of the actual system $\beta=0$.}
	\label{fig:pop-fake}
\end{figure}
We also plot the fictitious dynamics (i.e., $\mathrm{tr}(\rho_f(t)\sigma_z)$) in Fig.~\ref{fig:pop-fake} to demonstrate the freezing effect: with $\beta>0$, the population on $\ket{\uparrow}$ in the fictitious system has slower transfer than the actual system. A large enough $\beta$ (0.8) even completely freezes the spin in its initial state ($\ket{\uparrow}$). On the other hand, with $\beta<0$, the transfer from $\ket{\uparrow}$ to $\ket{\downarrow}$ is accelerated and the reverse transfer is obstructed. These results are consistent with the analysis in Sec.~\ref{sec:theory}.

\begin{figure}
	\centering
	\includegraphics{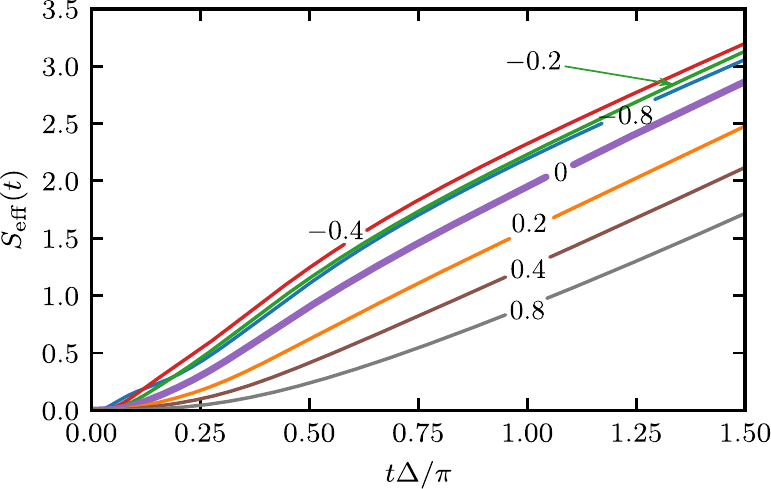}
	\caption{The entanglement growth dynamics for different $\beta$'s. The effective entanglement $S_\mathrm{eff}$ is a function of the von Neumann entanglements $\{S_n\}$ on every bond of a matrix products state: $S_\mathrm{eff} = \ln(\frac{1}{L-1}\sum_n e^{3S_n})^{1/3}$ where $L$ is the number of bonds \cite{rams2020breaking}. We use the binary logarithm to calculate von Neumann entanglement entropy.}
	\label{fig:entangle}
\end{figure}
Fig.~\ref{fig:entangle} shows that the members in the Hamiltonian family $\{\mathcal{H}(\beta)\}$ have different rates of entanglement growth. For positive $\beta$'s, the entanglement grows slower with time when $\beta$ is larger. For negative $\beta$'s, with a more negative $\beta$  (e.g., $-0.2\to-0.4$), the entanglement growth of the fictitious systems is faster, but such acceleration effect can reach its maximum at an critical value of $\beta$ (in this example, $\beta\sim-0.4$). If $\beta$ continues to become more negative (e.g., $\beta=-0.8$), the entanglement growth can be slower than its maximum speed ($\beta\sim-0.4$). This is because, with a very negative $\beta$, the population on $\ket{\uparrow}$ is transferred quickly to $\ket{\downarrow}$, then stays in $\ket{\downarrow}$, which disentangles the spin and the bath.

The similarity transformation present here suppresses the entanglement growth rate in the evolving quantum systems by almost 50\%. The singular values on the bonds (of the MPS) that are spatially close to the transformed site are affected by the transformation significantly: the singular values are localized on the first few eigenstates of the reduced density matrix. The bonds far from the transformed site are less affected. 
We can expect a more significant reduction of entanglement if a global transformation is applied to quantum many-body Hamiltonians. For example, for a coupled spin chain, a global similarity transformation can impact every spin. Consequently, after the transformation, each bond connecting neighboring  spins are dominated by a few relatively large singular values, reducing the entanglement in the MPS more significantly than single-spin systems. To show the advantage of the global transformation for multi-spin systems, we use a GHZ state of a 10-spin system as an example, and apply the transformation $e^{0.1\hat{\sigma}_z}$ to the first 0, 1, 2, \ldots, and 10 spins. The entanglement entropies (measured by the definition in the caption of Fig. \ref{fig:entangle}) of the states obtained with different number of transformations are 1.04,
1.01,
0.93,
0.82,
0.69,
0.57,
0.45,
0.36,
0.28,
0.22, and
0.17. A clear relationship between the reduced entanglement and the number of transformed spins can be observed.
The transformation is not limited to open quantum systems or spin systems, but may be useful for the simulations of general quantum many-body systems, such as the dynamics of electron-phonon models and correlated-electron models.

The similarity transformation $e^{\beta\hat{\sigma}_z}(\cdot)e^{-\beta\hat{\sigma}_z}$ has some alternatives. The generator of the transformation can be $\sigma^+$, $\sigma_-$, and any combinations of the sum and product of them. 
%Here, we give such an alternative and interpret the physical implications of the resulting Hamiltonian. 
% We use $e^{\hat{\sigma}_-}(\cdot)e^{-\hat{\sigma}_-}$ (i.e., $\beta=1$) to transform the Hamiltonian in Eq.~\eqref{eq:actual-Hamiltonian}. The result is
% \begin{align}
%     \underbrace{\Delta\begin{pmatrix}
%         -1 & 1\\
%         0  & 1
%     \end{pmatrix}}_\mathrm{spin} + \underbrace{\sum_n c_n\begin{pmatrix}
%         1 & 0\\
%         2  & -1
%     \end{pmatrix}\otimes(\hat{a}^\dagger_n + \hat{a}_n)}_\mathrm{spin-bath} + \sum_{n}\omega_n \hat{a}^\dagger_n\hat{a}_n.
%     \label{eq:alternative-hamiltonian}
% \end{align}
% The spin part in the above Hamiltonian is a biased two-level system with only the transition possibility from $\ket{\downarrow}$ to $\ket{\uparrow}$. In the spin-bath interaction term, the system coupling operator obtains an additional transition term from $\ket{\uparrow}$ to $\ket{\downarrow}$ which can be interpreted as the non-Condon effect. 
These alternative transformations demonstrate the great potential of the similarity-transformation method we present in this paper.

The entanglement suppression from the particular choice of the transformation $e^{\beta\hat{\sigma}_z}(\cdot)e^{-\beta\hat{\sigma}_z}$ 
used here arises from the suppressed transition between the two eigenstates of the system part of the system-bath interaction term $\hat{A}=\hat{\sigma}_z$ (see Eq.~\eqref{eq:transformed-spin}). The transformed wave function of the whole system (spin+bath) $e^{\beta\hat{\sigma}_z}\ket{\psi}(t)$ is less entangled because the spin (the system) tends to stay in one of the eigenstates of $\hat{A}=\hat{\sigma}_z$.
For general system-bath interaction interactions, for example, $\hat{A}=x\hat{\sigma_z}+z\hat{\sigma}_z$ where $x$ and $z$ are scalars, the transformation $e^{\beta\hat{A}}(\cdot)e^{-\beta \hat{A}}$ can be an initial choice of the optimal similarity transformation, since it “projects” (approximately) the state of the spin to one of the eigenstates of the coupling operator $\hat{A}=x\hat{\sigma_x}+z\hat{\sigma}_z$. Our numerical tests show that this choice of transformation ($e^{\beta A}$)is 
better than other combinations of $\hat{\sigma}_z$ and $\hat{\sigma}_x$.

For the choice of $\beta$, large $|\beta|$'s could bring large numerical errors in the simulation. A large $|\beta|$ requires smaller singular-value threshold to recover the original density matrix as in Eq.~\eqref{eq:density-matrix-relation}. An empirical value for the optimal $\beta$ is $\sim0.8$ which suppress the population on $\ket{\uparrow}$ to $\sim20\%$ of its original value.

\section{Conclusion}
By introducing a particular similarity transformation to an quantum system, we introduced a family of non-Hermitian Hamiltonians, and show numerically the different growth speeds of entanglement in the evolution of open quantum system dynamics for the members in the family. The similarity transformation serves as a valve for the transitions between the quantum states of the system, and a parameter ($\beta$) in the transformation controls how the transitions are diminished or enhanced. The quantum-state transition enhancement and diminution can be tuned so that the quantum state of the system and bath is close to a disentangled state during the evolution, suppressing the growth of entanglement of matrix product states. Recent studies show that the dynamics from the $\mathcal{PT}$ and anti-$\mathcal{PT}$ symmetric Hamiltonians can produce reduced entanglement in the evolution of the system-bath dynamics \cite{cen2022anti,gardas2016pt}. Our result is consistent with the findings in these studies.
The similarity transformation introduced in this letter is simple, and it produced pronounced effect on the growth speed of entanglement even for a single-spin Hamiltonian. The numerical advantages can be further manifested in the simulation of non-Markovian dynamics of large multi-state systems where high entanglement prohibits long-time simulations \cite{rams2020breaking,lacroix2021unveiling}. The slow growth of entanglement is favorable to the MPS simulation of, but not limited to, open quantum systems, as the similarity transformation is general.

Future studies can explore other forms of similarity transformations that can further slow down the growth of entanglement. 
The combination of the similarity-transformation method and the basis-transformation methods (e.g., the chain transformations) \cite{chin2010exact,liu2021improved,rams2020breaking,de2015discretize} is also possible, to change the growth speed even further.
It is also of interest to investigate the effect of similarity transformations on other exact and approximate numerical methods, including the hierarchical equation of motion, quasi-adiabatic path integral, and quantum master equations.

\begin{acknowledgments}
	This material is based upon work supported by the National Science Foundation under Grant No. 1955138. We thank Jonathon L. Yuly for stimulating discussions.
\end{acknowledgments}

%\section*{References}
%\nocite{*}
\bibliography{references}% Produces the bibliography via BibTeX.

\end{document}